\documentclass[conference]{IEEEtran}
\IEEEoverridecommandlockouts
\usepackage{cite}
\usepackage{amsmath,amssymb,amsfonts}
\usepackage{graphicx}
\usepackage{textcomp}
\usepackage[letterpaper, top=54pt, bottom=54pt, left=54pt, right=54pt]{geometry}
\usepackage{xcolor}
\usepackage[T1]{fontenc}
\usepackage{hyperref}
\usepackage{tabularx}
\usepackage{booktabs,threeparttable,makecell}
\usepackage{pifont}
\usepackage{multirow}
\usepackage{rotating}
\usepackage{algorithm}
\usepackage{algpseudocode}
\usepackage{tikz}
\usepackage{comment}
\usepackage{amsmath}
\usepackage{array,makecell}
\newcolumntype{Y}{>{\raggedright\arraybackslash}X}
\newcommand{\Part}{\makecell[c]{Part.}}

\renewcommand{\arraystretch}{0.95}
\makeatletter
\newdimen\saved@topmargin
\newdimen\saved@textheight
\AtBeginDocument{%
	\saved@topmargin=\topmargin
	\saved@textheight=\textheight
	\addtolength{\topmargin}{18pt}%
	\addtolength{\textheight}{-18pt}%
	\global\@colht=\textheight
	\global\@colroom=\textheight
	\global\vsize=\textheight
}
\AddToHook{shipout/firstpage}{%
	\global\topmargin=\saved@topmargin
	\global\textheight=\saved@textheight
	\global\@colht=\textheight
	\global\@colroom=\textheight
	\global\vsize=\textheight
}
\makeatother

\def\BibTeX{{\rm B\kern-.05em{\sc i\kern-.025em b}\kern-.08em
    T\kern-.1667em\lower.7ex\hbox{E}\kern-.125emX}}
\begin{document}

\title{DRIFT: Driving Risk Inference via Field Transmission for Human-like Autonomous Driving
\thanks{© 2026 IEEE.  Personal use of this material is permitted.  Permission from IEEE must be obtained for all other uses, in any current or future media, including reprinting/republishing this material for advertising or promotional purposes, creating new collective works, for resale or redistribution to servers or lists, or reuse of any copyrighted component of this work in other works.}
\thanks{This paper has been accepted by IEEE International Conference on Intelligent Transportation Systems 2026.}
}

\author{Zian Wang, Yiming Shu, Zejian Deng, Chen Sun \IEEEmembership{Member, ~IEEE}
\thanks{All authors are with the Department of Data and Systems Engineering, The University of Hong Kong, Hong Kong SAR, China.}
\thanks{Corresponding author: Chen Sun (c87sun@hku.hk).}
}

\maketitle

\begin{abstract}
Risk fields offer spatially structured alternatives to scalar safety metrics. However, hand-crafted static risk field models struggle with occlusion and topology-driven propagation. We present \textbf{DRIFT}, a spatiotemporal risk field governed by an advection–diffusion–reaction partial differential equation (PDE), with an optional telegrapher term. DRIFT draws on three sources: anisotropic Gaussian kernels to capture velocity-induced risk, occlusion-aware latent hazards behind large vehicles, and topology-coupled merge-zone conflict pressure. We further introduce field-centric evaluation metrics to complement the existing Surrogate Safety Measures (SSMs), including Lane-Change Risk Differential, Temporal Anticipation Index, Occlusion Sensitivity Index, and Occlusion Response Latency. Experiments on real-world traffic datasets show that DRIFT reduces occlusion response latency by $52\%$, and lowers the near-collision rate under occlusion by $2.1\%$ compared with selected baselines in synthetic scenarios. The code is available at \url{https://github.com/SAS-HKU/DRIFT.git}.
\end{abstract}


\section{Introduction} \label{introduction}
\begin{figure*}[!t]
    \centering
    \includegraphics[width=0.9\textwidth]{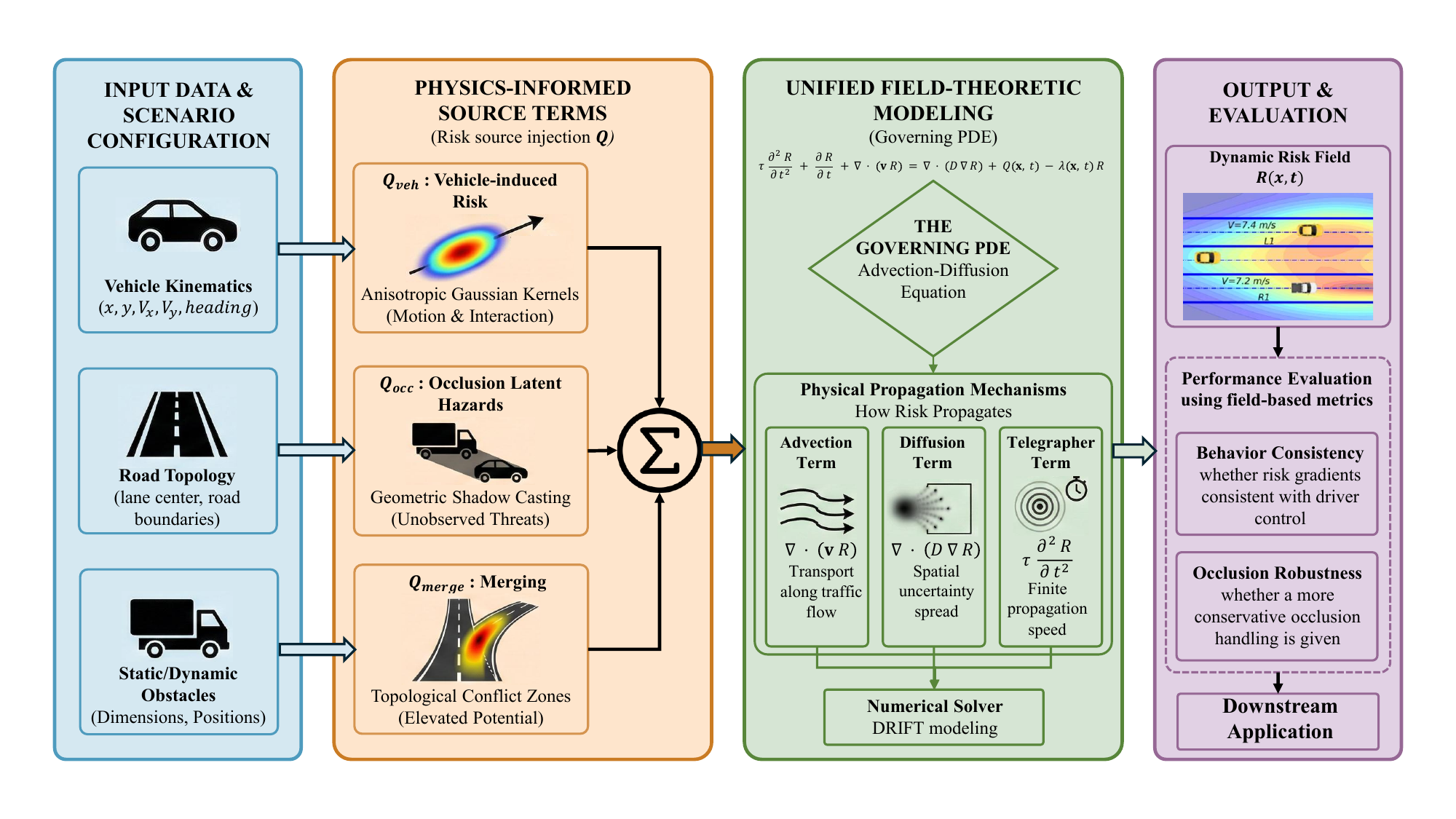}
    \caption{Overall structure of the PDE-based risk inference via field transmission framework for human-like autonomous driving.}
    \label{fig_methodology}
\end{figure*}

Understanding and modeling risks in mixed traffic environments remain
fundamental challenges for modern Intelligent Transportation Systems
(ITS), especially in interactive scenarios such as highway merging and
lane changing, where vehicles must reason about surrounding agents'
intentions, constraints, and uncertainties~\cite{cheng2025driving,
sun2023toward}. Traditional safety metrics provide useful signals, but
they are typically pairwise, instantaneous, and scalar, making it
difficult to capture the spatial structure of multi-agent interactions.

Field-based representations, also referred to as risk fields or
potential fields, offer spatially structured alternatives by defining
risk as a continuous scalar function over the Bird's Eye View (BEV) plane. Their local magnitudes and gradients
encode hazard distribution and spatial influence, enabling downstream
modules to reason beyond discrete agent locations
\cite{rasidescu2024artificial}. Existing risk-field methods include
hand-crafted static potentials based on kinematics or interaction
primitives~\cite{wang2024modeling}, physics-informed dynamic fields with explicit
transmission structure, and data-driven or hybrid risk maps that learn dense
risk/occupancy representations from trajectory data
\cite{jiang2024edrf,wang2025data,liu2026risknet}. However, occlusion and
partial observability remain critical challenges, and existing
evaluations still largely rely on surrogate safety measures or scenario
stress tests~\cite{tian2025driving}.

Despite this progress, three limitations remain. First, many risk fields
are static or frame-wise, and therefore lack a shared transmission law
for temporal anticipation. Second, occlusion is often handled through
masking or heuristic padding rather than explicit latent-risk
propagation and clearance. Third, field models are commonly evaluated
using scalar Surrogate Safety Measures (SSMs), such as
Time-to-Collision (TTC),
Post-Encroachment-Time (PET)~\cite{hou2024equipping}, or downstream
collision statistics~\cite{ma2025modeling}, which cannot fully diagnose
whether the spatial structure and propagation behavior of a risk field
are correct, especially in occluded or far-ahead regions.

To address these gaps, we introduce \textbf{DRIFT}: \textbf{D}riving \textbf{R}isk \textbf{I}nference via \textbf{F}ield \textbf{T}ransmission governed
by a partial differential equation (PDE)-based field transmission model, with structure depicted in
Fig.~\ref{fig_methodology}. Our contributions are as follows:
\begin{enumerate}
  \item A unified advection--diffusion--reaction PDE that provides a
  principled transmission substrate for risk, preserving temporal memory
  through decay, propagating hazards downstream through advection, and
  spreading spatial uncertainty through diffusion.

  \item A geometry-coupled occlusion decay mechanism in which the field's
  local decay rate responds directly to the rate of change of the shadow
  mask.

  \item A field-native evaluation protocol comprising behavioral,
  propagation, and occlusion-response metrics.
\end{enumerate}

The remainder of this paper is structured as follows.
Sections~\ref{sec:formulation} and~\ref{sec:methodology} formulate the
problem and methodology. Section~\ref{sec:experiment_results} presents
experiments, and Section~\ref{sec:conclusion} concludes the paper.

\section{Preliminaries}
\label{sec:formulation}
\subsection{Modeling Risk Potential Field}
\begin{figure}
    \centering
    \includegraphics[width=0.45\textwidth]{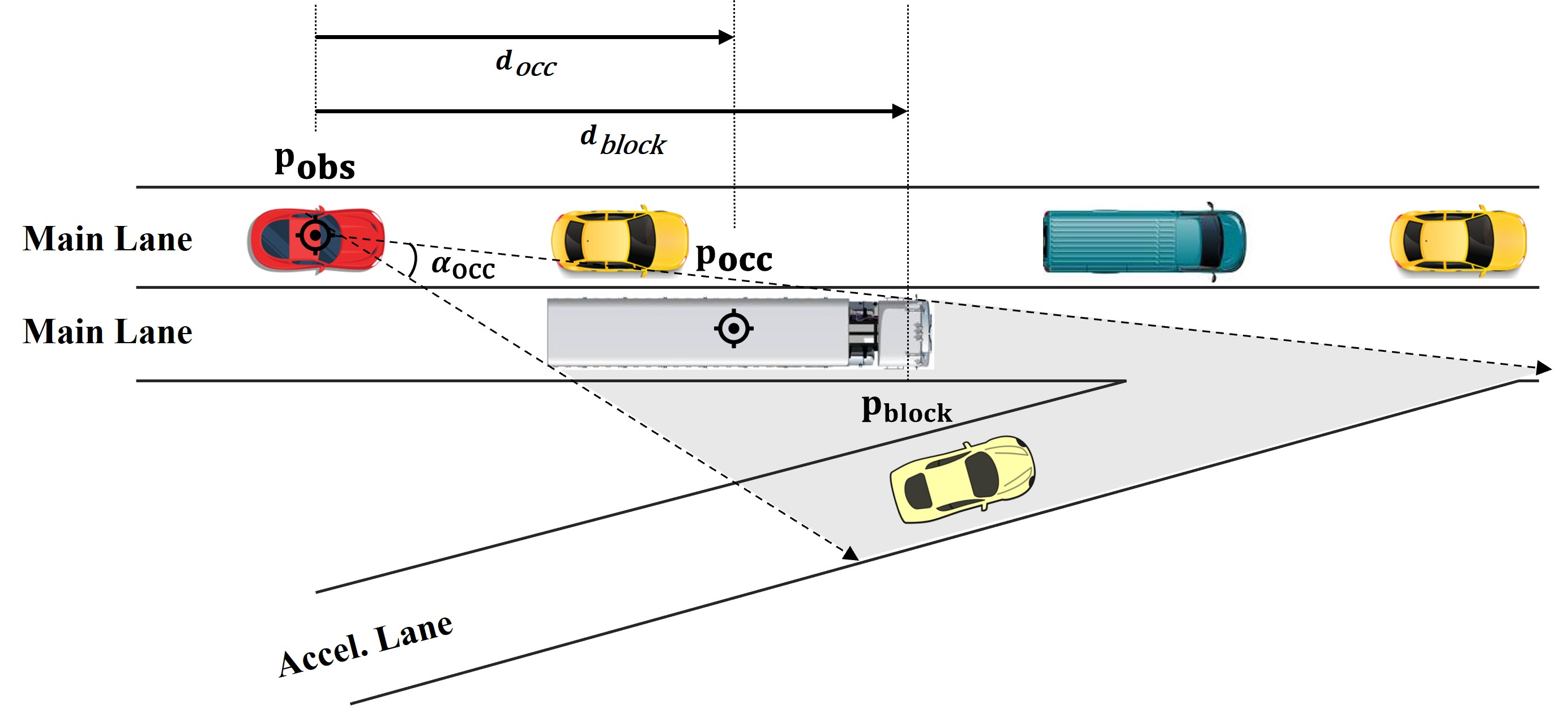}
\caption{Occlusion geometry in highway merging. The ego vehicle (observer) in the Main Lane with the position $\mathbf{p}_{\text{obs}}$ in the main lane cannot perceive the blocked vehicle with position $\mathbf{p}_{\text{block}}$ in the Acceleration (Accel.) Lane due to the heavy truck (occluder) with the position $\mathbf{p}_{\text{occ}}$. 
}
\label{fig:occlusion_geometry}
\end{figure}

We adopt a potential-field view in which risk is represented as a non-negative scalar potential over the BEV plane, where higher values indicate regions that should be avoided or treated conservatively by a downstream policy \cite{xie2022distributed}.
In DRIFT, the field is defined in the ego-centric perception frame: $\Omega \subset \mathbb{R}^2$ denotes the spatial domain within the ego vehicle's sensor range, and $t \in [0,T]$.
Formally, the risk field is denoted as:
$R(\mathbf{x}, t): \Omega \times [0,T] \rightarrow \mathbb{R}^+$, where $\mathbf{x}=(x,y)\in\mathbb{R}^2$ is a 2D spatial position vector in the BEV plane. At each time $t$, the scene contains the ego vehicle and surrounding agents indexed by $i$, each with state $(\mathbf{x}_i(t), \mathbf{v}_i(t), \mathbf{a}_i(t))$ expressed in the same BEV coordinates.
Occlusion is modeled from the ego viewpoint via a time-varying shadow region $\Omega_{\text{occ}}(t)$ induced by large vehicles (e.g., truck--trailers), and a smoothed shadow mask $S(\mathbf{x}, t)\in[0,1]$ for stable coupling to the field dynamics (Fig.~\ref{fig:occlusion_geometry}). 
\subsection{Dataset Loading and Field Construction}
We construct the time-indexed source term $Q(\mathbf{x},t)$ and the corresponding field $R(\mathbf{x},t)$ from raw trajectory logs.
For BEV datasets, we parse the CSV tracks to obtain, for each frame, the ego and surrounding-agent states $(\mathbf{x}_i(t), \mathbf{v}_i(t), \mathbf{a}_i(t))$ in a common BEV coordinate system.
These states are then converted into three additive sources:
(i) a vehicle-interaction source $Q_{\text{veh}}$ obtained by placing
anisotropic Gaussian kernels at each agent; (ii) an occlusion source
$Q_{\text{occ}}$ obtained by computing the ego-view shadow mask
$S(\mathbf{x},t)$ and injecting latent hazard within the corresponding
shadow region $\Omega_{\text{occ}}(t)\subset\Omega$; and (iii) a
topology/merge source $Q_{\text{merge}}$ derived from lane geometry and
conflict-zone priors.
The full source is $Q = Q_{\text{veh}}+Q_{\text{occ}}+Q_{\text{merge}}$, which is evaluated on a fixed BEV grid per frame and used to evolve the spatiotemporal field $R(\mathbf{x},t)$ via the PDE transmission model described in Section~\ref{sec:methodology}.

\section{Methodology}
\label{sec:methodology}
\subsection{Risk Field Transmission Model}
\label{sec:field_modeling}

Given the source field $Q(\mathbf{x},t)$ constructed from the scene
representation in Section~\ref{sec:formulation}, DRIFT evolves the risk
potential $R(\mathbf{x},t)$ over the ego-centric BEV domain $\Omega$
using an advection--diffusion--reaction transmission model:
\begin{equation}
\frac{\partial R}{\partial t}
=
\underbrace{\nabla\cdot(D\nabla R)}_{\text{diffusion}}
-
\underbrace{\nabla\cdot(\mathbf{v}R)}_{\text{advection}}
+
\underbrace{Q(\mathbf{x},t)}_{\text{source}}
-
\underbrace{\lambda(\mathbf{x},t)R}_{\text{decay}}.
\label{eq:adr}
\end{equation}

Here, $\mathbf{v}(\mathbf{x},t)=\mathbf{v}_{\text{flow}}+
\mathbf{v}_{\text{topo}}$ is the effective transmission velocity,
$D(\mathbf{x},t)$ is the spatial diffusivity, and
$\lambda(\mathbf{x},t)$ is the decay rate. An optional inertial
telegrapher term $\tau\partial^2 R/\partial t^2$ can be added to the
left-hand side to enforce finite propagation speed; in our experiments,
we set $\tau=0$ for computational efficiency and numerical stability.

The total source field is the additive composition:
\begin{equation}
Q(\mathbf{x},t)
=
Q_{\text{veh}}(\mathbf{x},t)
+
Q_{\text{occ}}(\mathbf{x},t)
+
Q_{\text{merge}}(\mathbf{x},t),
\label{eq:source-sum}
\end{equation}
where:
\begin{align}
Q_{\text{veh}}(\mathbf{x},t)
&=
\sum_i w_i G(\mathbf{x};\boldsymbol{\mu}_i,\mathbf{P}_i),
\label{eq:qveh}
\\
Q_{\text{occ}}(\mathbf{x},t)
&=
\mathbf{1}_{\Omega_{\text{occ}}(t)}(\mathbf{x})
\,g(\mathbf{x},t)\,p_{\text{emerge}}(\mathbf{x},t),
\label{eq:qocc}
\\
Q_{\text{merge}}(\mathbf{x},t)
&=
k_{\mathrm{amb}}\rho_{\text{merge}}(\mathbf{x})
+
k_{\mathrm{veh}}\rho_{\text{merge}}(\mathbf{x})\rho_{\text{veh}}(\mathbf{x},t).
\label{eq:qmerge}
\end{align}

The vehicle term places anisotropic Gaussian kernels at observed agents, the occlusion term injects latent risk inside the ego-view shadow region, and the merge term encodes topology-driven conflict pressure near merge zones. These sources anchor the field to scene evidence, while the shared operator in \eqref{eq:adr} propagates, smooths, and clears risk over time. 

\subsection{Model Design Rationale}
\label{sec:drift_reflection}

\textbf{Role of each PDE term.}
The four components are complementary
rather than interchangeable.
The source term $Q(\mathbf{x},t)$ anchors the field to observable
scene evidence (vehicle interaction, occlusion, and topology
conflict).
Advection transmits injected hazards along the effective flow, allowing risk to emerge before an interaction reaches the ego vehicle.
Diffusion smooths sharp peaks and spreads positional uncertainty,
which is especially important inside occlusion regions where the
hidden-agent distribution is broad.
Decay removes stale hazard and prevents persistent wakes from
dominating the map after the causal interaction has ended or
visibility has returned.

\textbf{Initialization and boundary conditions.}
We initialize the field as $R(\mathbf{x},0)=0$ so that risk is
purely evidence-driven and does not inherit arbitrary priors.
At the sensor boundary $\partial\Omega$, we apply zero-flux
Neumann conditions $\partial R/\partial n|_{\partial\Omega}=0$ to
avoid numerical loss of hazard through the boundary of the
perceived domain.
Downstream, an absorbing sponge layer
($\lambda_{\text{sponge}} \propto \|\mathbf{x}-\mathbf{x}_s\|^2$) dissipates advected
risk as it exits the sensor range, preventing artificial
back-reflections that would otherwise re-contaminate upstream
estimates.

\textbf{Occlusion Handling}. The decay rate is coupled to the shadow mask \(S(x,t)\). When a shadow expands, decay is reduced so that latent risk can accumulate; when a shadow retreats, decay is increased to clear risk no longer supported by occlusion. We initialize
$R(\mathbf{x},0)=0$, use zero-flux boundary conditions at the sensor
boundary, and apply a downstream sponge layer to dissipate risk leaving
the observable domain.

\subsection{Occlusion-Aware Scenario Extraction}
\label{sec:occlusion-extraction}

To evaluate occlusion handling under a consistent protocol, we first
extract occlusion windows from the raw trajectory logs and then apply all
methods to the same selected intervals. Algorithm~\ref{alg:occlusion} details the occlusion-window extraction process and relevant parameters. Let $k$ denote the frame index, $t_k=k/\nu$ the corresponding time, and
$\mathbf{p}_j^k\in\mathbb{R}^2$ the BEV position of agent $j$ at frame
$k$. For the ego vehicle $e$, we transform each candidate heavy vehicle
$j$ into the ego body frame as:
\begin{equation}
    \mathbf{r}_{e,j}^k =
    \mathbf{R}(\psi_e^k)
    \left(\mathbf{p}_j^k-\mathbf{p}_e^k\right),
    \quad
    \mathbf{R}(\psi_e^k)=
    \begin{bmatrix}
    \cos\psi_e^k & \sin\psi_e^k \\
    -\sin\psi_e^k & \cos\psi_e^k
    \end{bmatrix},
\end{equation}
where $\mathbf{r}_{e,j}^k=(r_{\parallel,e,j}^k,r_{\perp,e,j}^k)^\top$
contains the longitudinal and lateral relative coordinates in the ego
frame. A frame is flagged as occlusion-relevant if at least one
heavy-class agent lies in the forward selection zone:
\begin{equation}
    \mathcal{Z}_{\text{front}}
    =
    \left\{
    \mathbf{r}=(r_{\parallel},r_{\perp}) :
    0 < r_{\parallel} < 90\,\mathrm{m},
    \ |r_{\perp}| < 18\,\mathrm{m}
    \right\}.
\end{equation}

The set of flagged frames is therefore:
\begin{equation}
    \mathcal{F}_{\text{occ}}
    =
    \left\{
    k :
    \exists j\neq e,\;
    c_j\in\mathcal{C}_{\text{heavy}},
    \ \mathbf{r}_{e,j}^k\in\mathcal{Z}_{\text{front}}
    \right\}.
\end{equation}

We group $\mathcal{F}_{\text{occ}}$ into maximal contiguous runs and pad
each run by $T_{\mathrm{pad}}=2\,\mathrm{s}$ on both sides. 

\noindent{\textbf{Connection to the PDE field variables.}}
The extraction procedure does not redefine the occlusion field
$\Omega_{\mathrm{occ}}(t)$ used in~\eqref{eq:qocc}. Instead, it
selects the temporal intervals in which such occlusion-aware field
modeling is evaluated. Specifically, a forward heavy-vehicle selection
zone $\mathcal{Z}_{\mathrm{front}}$ is used to identify occlusion-relevant
frames from raw trajectories. For each selected frame $t_k$, the
ego-view shadow mask $S(\mathbf{x},t_k)$ and the corresponding shadow
region
$\Omega_{\mathrm{occ}}(t_k)=\{\mathbf{x}\in\Omega:
S(\mathbf{x},t_k)>\theta_S\}$
are then computed on the same BEV grid as the risk field. 

\begin{algorithm}[t]
\caption{Occlusion-window extraction}
\label{alg:occlusion}
\begin{algorithmic}[1]
\Require Recording
$\mathcal{D}=\{(\mathbf{p}_j^k,\psi_j^k,c_j)\}_{j,k}$,
ego id $e$, frame rate $\nu$, padding time $T_{\mathrm{pad}}$,
heavy-class set $\mathcal{C}_{\mathrm{heavy}}$
\Ensure Occlusion-window set $\mathcal{W}_{\mathrm{occ}}$

\State $\mathcal{F}_{\mathrm{occ}}\gets\varnothing$
\State $\mathcal{Z}_{\mathrm{front}}
\gets
\{(r_{\parallel},r_{\perp}):
0<r_{\parallel}<L_{\parallel},\
|r_{\perp}|<L_{\perp}\}$

\ForAll{frames $k=1,\ldots,K$ in which ego $e$ is active}
    \State Obtain ego pose $(\mathbf{p}_e^k,\psi_e^k)$
    \State $\mathbf{R}_e^k \gets
    \begin{bmatrix}
    \cos\psi_e^k & \sin\psi_e^k\\
    -\sin\psi_e^k & \cos\psi_e^k
    \end{bmatrix}$

    \ForAll{agents $j\neq e$ active at frame $k$ with
    $c_j\in\mathcal{C}_{\mathrm{heavy}}$}
        \State $\mathbf{r}_{e,j}^k
        \gets
        \mathbf{R}_e^k(\mathbf{p}_j^k-\mathbf{p}_e^k)$

        \If{$\mathbf{r}_{e,j}^k\in\mathcal{Z}_{\mathrm{front}}$}
            \State $\mathcal{F}_{\mathrm{occ}}
            \gets
            \mathcal{F}_{\mathrm{occ}}\cup\{k\}$
            \State \textbf{break}
        \EndIf
    \EndFor
\EndFor

\State Group $\mathcal{F}_{\mathrm{occ}}$ into maximal contiguous runs
$\{[s_w,e_w]\}_{w=1}^{W}$
\State $\Delta_{\mathrm{pad}}\gets
\lceil T_{\mathrm{pad}}\nu\rceil$
\State \Return
$\mathcal{W}_{\mathrm{occ}}
=
\big\{
[\max(1,s_w-\Delta_{\mathrm{pad}}),
 \min(K,e_w+\Delta_{\mathrm{pad}})]
\big\}_{w=1}^{W}$
\end{algorithmic}
\end{algorithm}
For every $k\in\mathcal{W}_{\mathrm{occ}}$, we compute
$S(\mathbf{x},t_k)$, $\Omega_{\mathrm{occ}}(t_k)$,
$Q_{\mathrm{occ}}(\mathbf{x},t_k)$, and
$\lambda(\mathbf{x},t_k)$ on the same BEV grid used by \eqref{eq:adr}.
Here, $L_{\parallel}=90\,\mathrm{m}$ and $L_{\perp}=18\,\mathrm{m}$
define the ego-frame selection zone
$\mathcal{Z}_{\mathrm{front}}$ for extracting evaluation windows. 
$T_{\mathrm{pad}}=2\,\mathrm{s}$ preserves pre- and post-occlusion
field evolution, and $\mathcal{C}_{\mathrm{heavy}}$ denotes truck/trailer
occluder classes.
\subsection{Metrics and Evaluation Framework}
\label{sec:metrics_framework}

\textbf{Limitations of Traditional SSMs for Field Evaluation}
Traditional SSMs are insufficient for evaluating continuous risk fields because they reduce safety to pairwise, instantaneous scalars and discard spatial gradients, directional structure, and field coherence across the BEV domain.
They are also visibility-blind: TTC and related measures are undefined for occluded agents, so they cannot distinguish conservative shadow risk from risk collapse.
Finally, they are reactive under constant-velocity assumptions and therefore miss the lead-time benefit introduced by advection-based propagation.
We therefore introduce field-native metrics grounded in observable trajectory data and independent of the field model being assessed:

\noindent\textbf{LCRD} (Lane-Change Risk Differential) is the
fraction of recorded lane changes that move toward a lower-risk
cell:
\begin{equation}
  \mathrm{LCRD} =
    \tfrac{|\{i:R(\mathbf{x}_{\mathrm{curr},i})>R(\mathbf{x}_{\mathrm{target},i})\}|}
          {|\text{all lane changes}|}.
  \label{eq:lcrd}
\end{equation}

\noindent\textbf{TAI} (Temporal Anticipation Index) measures how
early the field threshold is crossed relative to the hazardous
event, normalized by human reaction time $t_r\!=\!1.5$\,s:
\begin{equation}
\mathrm{TAI} = \frac{1}{N}\sum_i \frac{t^i_{\mathrm{event}} - t^i_{\mathrm{field}}}{t_r}.
  \label{eq:tai}
\end{equation}

\noindent\textbf{RPR} (Risk Persistence Ratio) quantifies
post-event risk memory over a $\Delta T\!=\!2$\,s window:
\begin{equation}
  \mathrm{RPR} =
    \tfrac{\int_{t_e}^{t_e+\Delta T} R(\mathbf{x},t)\,dt}
          {R_{\mathrm{peak}}\cdot\Delta T}.
  \label{eq:rpr}
\end{equation}

\noindent\textbf{OSI}, \textbf{ORL}, $\Delta$\textbf{Coll.\%},
and \textbf{Temp.}\ characterize occlusion robustness.
OSI $= \Delta\bar{R}/\Delta\rho$ measures risk sensitivity to
the occlusion ratio $\rho$; ORL $= |t_{R>\theta} -
t_{\mathrm{shadow}}|$ measures shadow response latency;
$\Delta$Coll. counts the relative increase in near-collision
events under occlusion versus full visibility; and Temp.\ is
the frame-to-frame autocorrelation of region-mean risk.


\section{Experiments and Results}
\label{sec:experiment_results}

\subsection{Experimental Setup}

We evaluate on exiD~\cite{moers2022exid}, rounD~\cite{krajewski2020round}, and inD~\cite{bock2020ind} on a unified $150\!\times\!70$ BEV grid
at 20\,Hz.
Baselines include static physics-informed fields (TPF~\cite{wang2024modeling},
APF~\cite{xie2022distributed}, DPDRF~\cite{liu2025dpdrf})
and dynamic data-driven fields (EDRF~\cite{jiang2024edrf},
RiskNet~\cite{liu2026risknet}).
Selected PDE and source parameters are listed in
Table~\ref{tab:parameters}.
\begin{table}[ht]
\centering
\small
\caption{Selected PDE and source parameters.}
\label{tab:parameters}
\begin{tabularx}{\columnwidth}{@{} l l l >{\raggedright\arraybackslash}X @{}}
\toprule
\textbf{Symbol} & \textbf{Value} & \textbf{Unit} & \textbf{Description} \\
\midrule
\multicolumn{4}{l}{\textit{Source Parameters}} \\
$\sigma_x, \sigma_y$ & 12.0, 3.0 & m & Longitudinal/lateral kernel spread \\
$L_{\mathrm{prox}}$ & 50.0 & m & Distance attenuation length \\
$v_{\mathrm{ref}}$ & 5.0 & m/s & Relative-speed reference \\
$\beta_a, \beta_{\mathrm{brake}}$ & 0.3, 0.8 & --- & Acceleration/braking source weights \\
$a_{\mathrm{ref}}$ & 3.0 & m/s$^2$ & Acceleration normalization \\
$\gamma_a, t_{\mathrm{react}}$ & 0.5, 1.5 & ---, s & Braking kernel shift \\
\midrule
\multicolumn{4}{l}{\textit{PDE Parameters}} \\
$D_0, D_{\mathrm{occ}}$ & 1.0, 3.0 & m$^2$/s & Base and occlusion diffusion \\
$\lambda_0, \lambda_{\mathrm{sh}}$ & 0.15, 1.3 & 1/s & Base and shadow decay \\
$\lambda_g$ & 1.0 & 1/s & Geometry-coupled decay coefficient \\
\midrule
\multicolumn{4}{l}{\textit{Merge Parameters}} \\
$k_{\mathrm{amb}}, k_{\mathrm{veh}}$ & 0.08, 0.6 & --- & Ambient and vehicle-gated merge weights \\
\bottomrule
\end{tabularx}
\end{table}

\subsection{Behavior Consistency}
 We first evaluate
behavior consistency, which measures whether field structure
agrees with recorded driver decisions. We report LCRD and RPR for
all methods, since both are defined for static and dynamic fields. Table~\ref{tab:results} summarizes all results.
\begin{table*}[t]
  \centering
  \caption{Performance Comparison Across All Metrics.
           ``\textbf{--}'' denotes metrics that are structurally
           undefined for non-temporal (static) formulations.}
  \label{tab:results}
  \setlength{\tabcolsep}{5pt}
  \begin{tabular}{l cc cc cc cc}
    \toprule
    & \multicolumn{2}{c}{\textit{Behavior Consistency}}
    & \multicolumn{2}{c}{\textit{Propagation Advantage}}
    & \multicolumn{4}{c}{\textit{Occlusion Robustness}} \\
    \cmidrule(lr){2-3}\cmidrule(lr){4-5}\cmidrule(lr){6-9}
    \textbf{Method}
      & LCRD$\uparrow$ & RPR$\uparrow$
      & TAI$\uparrow$ & (TAI$\!\times\!t_r$)
      & OSI$\uparrow$ & ORL$\downarrow$
      & $\Delta$Coll.$\downarrow$ & Temp.$\uparrow$ \\
    \midrule
    \multicolumn{9}{l}{\textit{Physics-Informed (Static)}} \\
    TPF~\cite{wang2024modeling}
      & 62\% & 0.38
      & \multicolumn{2}{c}{-- $^{\dagger}$}
      & 0.08 & -- & +9.3\% & 0.61 \\
    GVF~\cite{zhang2021spatiotemporal}
      & 71\% & 0.42
      & \multicolumn{2}{c}{-- $^{\dagger}$}
      & 0.12 & -- & +8.1\% & 0.72 \\
    DPDRF~\cite{liu2025dpdrf}
      & 68\% & 0.41
      & \multicolumn{2}{c}{-- $^{\dagger}$}
      & 0.15 & -- & +6.8\% & 0.68 \\
    \midrule
    \multicolumn{9}{l}{\textit{Dynamic (Data-Driven)}} \\
    EDRF~\cite{jiang2024edrf}
      & 78\% & 0.61 & 0.18 & ($+$0.27\,s)
      & 0.28 & 0.82\,s & +4.7\% & 0.81 \\
    RiskNet~\cite{liu2026risknet}
      & 82\% & 0.67 & 0.24 & ($+$0.36\,s)
      & 0.35 & 0.65\,s & +3.5\% & 0.85 \\
    \midrule
    \textbf{DRIFT (Ours)}
      & \textbf{89\%} & \textbf{0.74}
      & \textbf{0.41} & \textbf{($+$0.62\,s)}
      & \textbf{0.52} & \textbf{0.31\,s}
      & \textbf{+1.4\%} & \textbf{0.91} \\
    \bottomrule
  \end{tabular}
  \vspace{2pt}

  \noindent\footnotesize
  $^{\dagger}$~TAI and ORL are undefined for static (frame-wise)
  risk fields because these methods recompute risk from
  instantaneous states and carry no temporal memory.
  Reporting TAI~$\approx 0$ for static methods would be
  misleading, as it conflates a structural impossibility with
  a measurable outcome.
\end{table*}
DRIFT achieves the best overall alignment with observed driver
actions. It attains an LCRD of 89\% versus 82\% (RiskNet) and 78\%
(EDRF), indicating that recorded lane changes are more often
directed toward lower-risk regions. DRIFT also yields
the highest post-event risk retention (RPR 0.74), suggesting that
it preserves salient interaction risk without excessive
lingering.

\subsection{Propagation Advantage}
Since
static fields are recomputed frame-by-frame and carry no temporal
memory, TAI and ORL are structurally undefined for static
baselines; we therefore report these metrics only for dynamic
methods. DRIFT provides the largest anticipatory
margin. It reaches TAI $=0.41$ (corresponding to $\approx 0.62$\,s
at $t_r=1.5$\,s), compared with 0.24 for RiskNet and 0.18 for
EDRF. This improvement reflects PDE-driven transmission that
carries injected hazard downstream as a structured wake, rather
than reacting only at the onset of conflict.

\subsection{Occlusion Robustness}
Occlusion-robustness metrics evaluate whether the field remains conservative inside shadows while clearing latent risk promptly once visibility returns. We report OSI, ORL, $\Delta$Coll., and Temp.\ to cover sensitivity, response latency, safety impact, and temporal stability.

\begin{figure*}[h]
  \centering
  \includegraphics[width=0.8\textwidth]{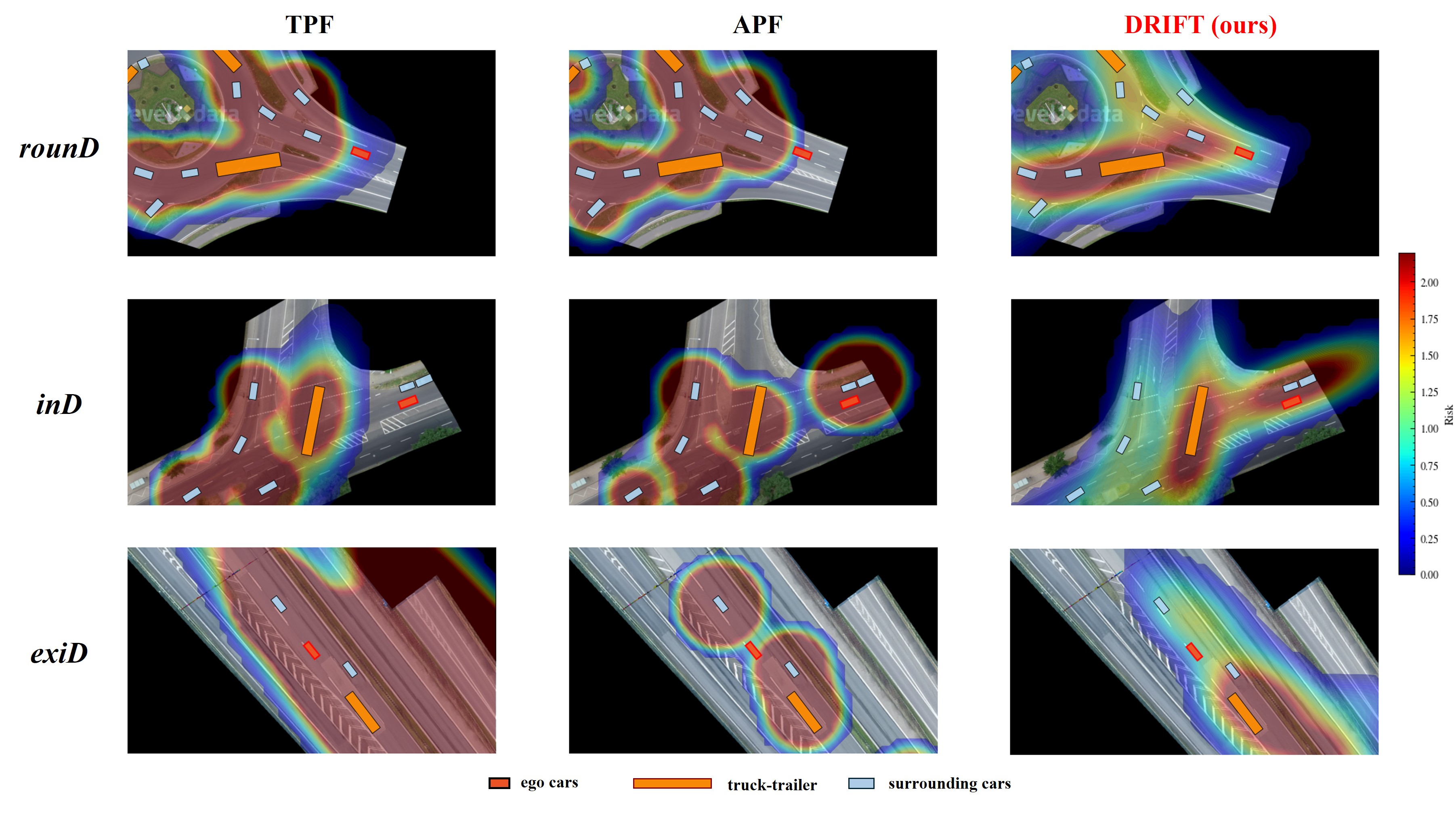}
  \caption{Risk field modeling on historical ego trajectories recorded in various traffic scenarios.}
  \label{fig:drift_benchmark}
  
\end{figure*}

DRIFT produces the most conservative yet responsive field
evolution under occlusion. It achieves the highest occlusion
sensitivity (OSI 0.52) and the fastest occlusion response (ORL
0.31\,s), while reducing the near-collision rate increase under
occlusion to +1.4\% (vs.\ +3.5\% for RiskNet and +4.7\% for EDRF).
DRIFT also attains the strongest temporal stability (Temp.\ 0.91),
consistent with rapid clearance of latent risk once visibility is
restored via geometry-coupled decay.

\noindent\textbf{Note on static baselines.}
TAI and ORL are shown as “–” for static methods, namely TPF, APF, and DPDRF, because these metrics are structurally undefined for frame-wise fields with no temporal memory.
Fig.~\ref{fig:drift_benchmark} qualitatively compares DRIFT with TPF and
APF on rounD roundabout, inD intersection, and exiD highway-ramp
scenarios. The static baselines tend to assign broad and nearly uniform
risk around visible objects or road geometry, which can activate
redundant regions that are not directly involved in the current vehicle
interaction. In contrast, DRIFT treats risk as propagated information:
heavy truck-trailer occlusions induce latent risk that is transmited
along plausible traffic-flow and conflict directions rather than being
spread as a static potential. 

\subsection{Ablation Study}
\begin{figure*}
  \centering
  \includegraphics[width=\textwidth]{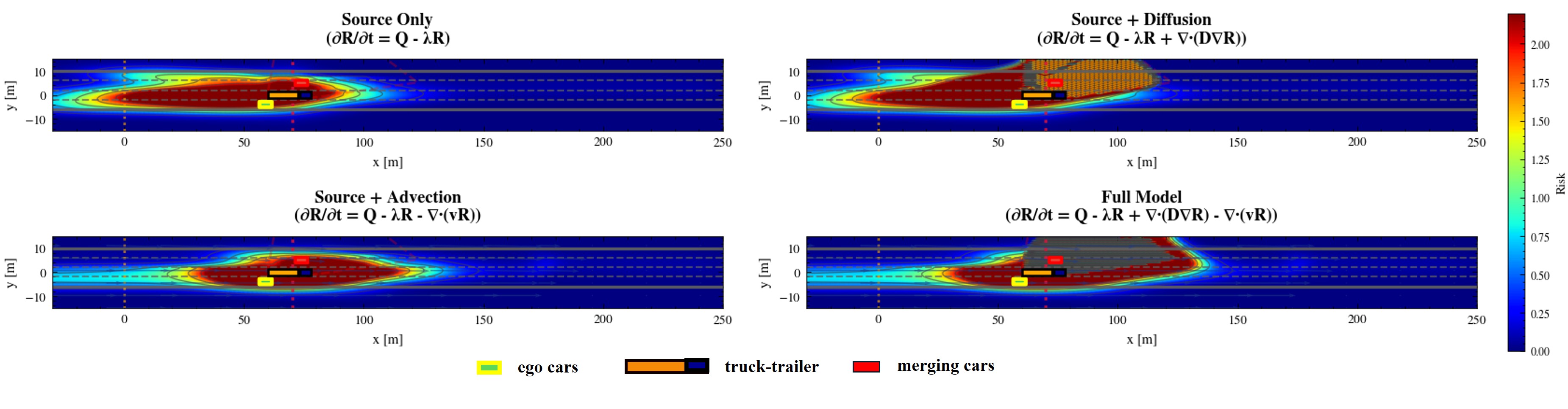}
  \caption{Risk field snapshots for four ablation settings on the same highway merge scene: source-only, +diffusion, +advection, and full DRIFT (+geometry-coupled decay).}
  \label{fig:ablation}
\end{figure*}
In Table~\ref{tab:ablation},  we ablate individual PDE terms.
Adding advection raises TAI from 0.12 to 0.35 ($+0.23$),
confirming that downstream transmission drives temporal anticipation.
Diffusion improves occlusion conservatism (OSI $0.35\!\to\!0.42$)
by spreading latent hazard energy beyond the geometric shadow
boundary.
Geometry-coupled decay (full model) is the primary driver of fast
visibility response (ORL $0.62\!\to\!0.31$\,s) and temporal
consistency ($0.85\!\to\!0.91$).

\begin{table}[h]
  \centering
  \caption{Ablation: PDE Term Contributions}
  \label{tab:ablation}
  \begin{tabular}{l cccc}
    \toprule
    \textbf{Config.} & TAI$\uparrow$ & RPR$\uparrow$ & OSI$\uparrow$ & ORL$\downarrow$ \;\; Temp.$\uparrow$ \\
    \midrule
    Source only        & 0.08 & 0.51 & 0.35 & 0.95\,s \quad 0.65 \\
    $+$ Diffusion      & 0.12 & 0.58 & 0.42 & 0.78\,s \quad 0.72 \\
    $+$ Advection      & 0.35 & 0.66 & 0.44 & 0.62\,s \quad 0.85 \\
    $+$ Decay (full)   & \textbf{0.41}
                       & \textbf{0.74} & \textbf{0.52}
                       & \textbf{0.31\,s} \quad \textbf{0.91} \\
    \bottomrule
  \end{tabular}
\end{table}

The risk field visualizations in Fig.~\ref{fig:ablation} complements Table~\ref{tab:ablation}
by visualizing the spatial effect of each PDE term on the same
scenario snapshot.
With source terms alone (upper left), risk is sharply localized
around each vehicle and conflict zone with no spatial spread.
Adding diffusion (upper right) produces smooth halos that extend
uncertainty into adjacent cells, raising OSI but introducing no
directional structure.
Adding advection without decay (lower left) creates a clear
downstream ``wake'' pattern in which risk trails behind
interacting vehicles along the traffic flow direction, explaining
the large TAI jump ($0.12\!\to\!0.35$) but also producing
residual risk that lingers after agents depart.
The full model with geometry-coupled decay (lower right) retains
the directional wake while actively clearing stale hazard,
yielding the highest temporal consistency (0.91) and the
fastest occlusion response (ORL $0.31$\,s).

\section{Conclusion} \label{sec:conclusion}
We presented DRIFT, a physics-informed spatiotemporal field modeling framework that explicitly propagates risk governed by a PDE transmission model. This formulation yields interpretable spatial risk structure and stable temporal dynamics, while enabling fast risk clearance when visibility is restored through geometry-coupled decay. The parameters of the PDE formulation remain tunable. Experiments on occlusion/merge scenarios from BEV datasets show that DRIFT better aligns field gradients with recorded driver actions and provides earlier warning compared to static fields and learned risk maps in the literature, while remaining robust under severe occlusion. Future work will extend DRIFT to a universal risk modeling that handles richer intersection geometries, more complex traffic scenarios, and multi-agent motion planning, and to integrate learned components such as uncertainty-aware sources and map priors. 

\section*{Acknowledgment}
The authors would like to acknowledge the financial support provided by the University Grants Committee of Hong Kong through the Early Career Scheme (Grant No. GRF/ECS-27206525).

\bibliographystyle{IEEEtran}
\bibliography{IEEEabrv, sources}

\end{document}